# Magnetotransport and Berry phase Tuning in Gd-doped Bi$_2$Se$_3$ Topological Insulator Single Crystals

Lei Chen[1], Shuang-Shuang Li[2], Weiyao Zhao[3,4], Abdulhakim Bake[4,5], David Cortie[4,5], Xiaolin Wang[4,5], Julie Karel[3], Han Li[†,1], and Ren-Kui Zheng[1,*]

[1]School of Physics and Materials Science, Guangzhou University, Guangzhou 510006, China

[2]School of Materials Science and Engineering and Jiangxi Engineering Laboratory for Advanced Functional Thin Films, Nanchang University, Nanchang 330031, P. R. China

[3]Department of Materials Science & Engineering, and ARC Centre of Excellence in Future Low-Energy Electronics Technologies, Monash University, Clayton VIC 3800, Australia

[4]ISEM, University of Wollongong, North Wollongong NSW 2500, Australia

[5]ARC Centre of Excellence in Future Low-Energy Electronics Technologies, University of Wollongong, North Wollongong NSW 2500, Australia

The Berry phase is an important concept in solids, correlated to the band topology, axion electrodynamics and potential applications of topological materials. Here, we investigate the magnetotransport and Berry phase of rare earth element Gd doped Bi$_2$Se$_3$ (Gd:Bi$_2$Se$_3$) topological insulator at low temperatures and high magnetic fields. Gd:Bi$_2$Se$_3$ single crystals show Shubnikov-de Haas (SdH) oscillations with nontrivial Berry phase while Bi$_2$Se$_3$ single crystals show zero Berry phase in SdH oscillations. The temperature dependent magnetization curves can be well fitted with the Curie-Weiss law in 3–300 K region, indicating no magnetic ordering in Gd:Bi$_2$Se$_3$ crystals. Moreover, Gd doping has limited influence on the quantum oscillation parameters (e.g., frequency of oscillations, the area of the Fermi surface, effective electron mass, Fermi wave vectors etc.), but has an impact on the Hall mobility, carrier density,

E-mail: *zrk@ustc.edu; †lihan@gzhu.edu.cn

and band topology. Our results demonstrate that Gd doping can tune the Berry phase of topological insulators effectively, which may pave a way for the future realization of many predicted exotic transport phenomena of topological origin.

## I. Introduction

The Berry phase, an important concept in many fields of physics, was proposed by Berry in 1984 [1]. The Berry phase is a Hamiltonian in a parameter space that can acquire a geometrical phase after undergoing a closed trajectory in adiabatic changes. This geometrical phase is only determined by the properties related to closed trajectory in a parameter space. In a crystal, Zak [2] argued that the Berry phase can be obtained on a moving electron in the Brillouin zone (as parameter space). As a result, the Berry phase of electrons of metals appears in the semiclassical quantization condition for its energy levels (Landau levels), which thus affects the Landau-level-related physical phenomena, e.g., Shubnikov-de Haas oscillations and de Haas-van Alphen oscillations [3,4]. The non-zero Berry phase correction to Landau quantization happens on electron orbits that link to band-contact lines, which demonstrates the Bloch band topology in solids [3]. The non-zero (nontrivial) Berry phase has been reported in Dirac semimetals [5-8], Weyl semimetals [9-11], topological nodal-line semimetals [12-18], and other topological materials [19-22]. Since the Berry phase is related to the band topology, the experimental observation is different in various materials: e.g., ~$0.75\pi$ and $\pi$ in graphite and graphene [23,24], $\pi$ in Rashba semiconductor BiTeI [25], 0.25-$0.75\pi$ in $Cd_3As_2$ [7,8].

The quantum oscillation behaviors are normally observed in metallic materials. However, recent reports broaden this knowledge in various insulating materials with special bulk-band symmetries and metallic surface states, e.g., $SmB_6$ [26,27], $YbB_{12}$ [28], Te [29], (V,

Sn):Bi$_{1.1}$Sb$_{0.9}$Te$_2$S [30]. The most discussed materials family is the three-dimensional (3D) topological insulators (TIs), which possess insulating bulk band and time-reversal-symmetry-protected Dirac surface state which was theoretically predicted in Bi$_2$Se$_3$ family [31] and experimental verified [32] in the last decade. The symmetry-protected surface state offers tremendous opportunities for spintronics, non-Abelian quantum computing, and energy-efficient electronic devices. Due to the Dirac band nature of the 3D TI's surface state, one may obtain the $\pi$ Berry phase shift on their Landau quantization related phenomena. However, in practice, the Berry phase situation for a 3D TI can be very complicated, for example, defects shift bulk Fermi level into conduction band in Bi$_2$Se$_3$ single crystals, resulting in non-Dirac bulk-dominant quantum oscillations with zero Berry phase [33]. Even for surface dominant transport of 3D TIs, the deviation of the surface state's dispersion from ideal linearity results in nontrivial Berry phase between zero and $\pi$, e.g., 0.44$\pi$ in Bi$_2$Te$_2$Se [34,35].

Another effective tunning of the Berry phase in a 3D TIs is the magnetic ion doping. Upon doping, the spin texture of topological surface state changes to a hedgehog-like spin texture, e.g., in a magnetic ion doped Bi$_2$Se$_3$, the surface state was gapped due to the ferromagnetic ordering, and the spin texture was turned to hedgehog like [36]. The Berry phase in such a magnetic ion doped topological insulator is defined by the spin texture of the Fermi surface, which, in this situation can be tuned from $\pi$ to zero via shifting the Fermi level to the Dirac point [36,37]. The ferromagnetic state in a topological insulator is very useful, which enables the novel quantum anomalous Hall effect [38], and paves the way to dissipationless electronic conductance in zero magnetic field. Moreover, tuning the Berry phase to zero in a ferromagnetic TI also provides the condition for axion electrodynamics [36,39]. The ferromagnetic ordering has been reported in 3$d$ transition metal doped topological insulators, e.g., Mn:Bi$_2$Te$_3$ [40,41], Cr:TlBiTe$_2$ [42], and Cr:Bi$_2$Se$_3$ [43]. Recently, rare earth elements are also considered as magnetic dopants in TI materials, which provides ferromagnetism without significantly harming the high mobility [44,45]. Previous researches show that Gd is a good dopant in 3D TIs, for instance, Gd:Bi$_2$Te$_3$ shows antiferromagnetism [46], Gd:TlBiTe$_2$ and Gd:Bi$_2$Se$_3$ show ferromagnetism with large Gd moments (5-7 $\mu_B$/Gd) [47,48]. Therefore, studying the magnetotransport properties of Gd doped topological insulators are helpful to understand the Berry phase. Here, we introduce a comprehensive magnetotransport study of Gd:Bi$_2$Se$_3$ single crystals to illustrate the nontrivial Berry phase in Gd doped TIs.

## II. Experiments

A modified Bridgeman method was employed to grow the topological insulator single crystals (Bi$_2$Se$_3$ and Gd:Bi$_2$Se$_3$). Briefly, high-purity Gd (99.9%), Bi (99.99%), and Se (99.99%) powders (~10 g) in the stoichiometric ratio were mixed and sealed in a quartz tube as starting materials. For the Gd doped sample, the starting nominate ratio is Gd$_{0.1}$Bi$_{1.9}$Se$_3$. The crystal growth was carried out using the following procedure: i) Heating the mixed powders to 1100 °C in 1 °C/min to completely melt them; ii) Maintaining at this temperature for 24 h to ensure uniform mixture of Gd, Bi, and Te atoms; and iii) Slowly cooling down to 500 °C at 2 °C/h to crystallize the sample; iv) naturally cooling to room temperature. Since the Gd dopants possess a high melting point, we set the melting temperature at 1100 °C in step i) and ii) to ensure full interaction among the molten elements in the liquid state. After the growth process, single-crystal flakes with a typical size of 5 × 5 × 0.2 mm$^3$ can be mechanically exfoliated from the ingot. The single crystals prefer to naturally cleave along the (001) planes, resulting in $c$-axis being the normal direction of the flakes as is commonly the case in this family of materials. The dopant distribution of Gd:Bi$_2$Se$_3$ flakes was confirmed using X-ray energy dispersive spectroscopy (EDS) coupled to a scanning electron microscopy (SEM). In the present work, we employ the EDS results as the final composition, which is determined to be Gd$_{0.02}$Bi$_{1.98}$Se$_3$.

The electronic transport properties were measured using a physical properties measurement system (PPMS, Dynacool-14T, Quantum Design). Hall measurements were performed on a freshly cleaved *ab* plane, using room-temperature cured silver paste. The electric current was parallel to the *ab* plane while the magnetic field was perpendicular to the *ab* plane. The angle dependence of the magnetoresistance (MR) was measured using a horizontal rotational rig mounted on the PPMS. Before rotation, the sample alignment was designed to make sure that the magnetic field was always perpendicular to the electric current. The magnetic measurements were conducted using the vibration sample magnetometer (VSM) equipped on PPMS. During temperature dependent magnetization (MT) measurements, the samples were cooled to 3 K with 500 Oe magnetic field (FC mode) and zero magnetic field (ZFC mode), respectively, after which the magnetization data were collected in the heating process, with applied magnetic field of 500 Oe. Magnetic hysteresis (MH) curves were obtained by scanning the magnetic field between 5 T and -5 T at certain temperatures.

## III. Results and Discussion

Before we conduct electronic transport measurements, EDS measurements were performed on freshly cleaved surface of crystals to analyze the distribution of Gd dopants. A relatively large area (500×670 μm$^2$) was selected to conduct the measurements to ensure the accuracy of element ratio in the crystal, as shown in Fig. 1(a). The elemental mapping was superimposed to SEM images to illustrate the elemental distribution with surface morphology. There're several wrinkles on the flat surface of the crystal, as demonstrated by the SEM image, which, however are not observed in elemental mapping image. The wrinkles are quite common on a cleaved quasi-two-dimensional crystal's surfaces. The SEM backscattered image and elemental mapping results suggest that the Gd dopant is uniformly distributed in the $Bi_2Se_3$ crystal without segregation. Further, we employ EDS to estimate the Gd doping level in the as-grown single crystals. Fig. 1(b) shows the characteristic peaks of all elements, in which the peak intensity related to the Bi and Se elements are much stronger than that of Gd element. After analyzing the peak intensity via the AZtec software, the elemental concentrations of two different Gd:$Bi_2Se_3$ samples are summarized in Table 1. The selenium vacancies are observed in both samples, which agrees with the literatures [33,49], and is the main reason for the Fermi level shifting into the conduction band. Gd doping level in both samples are stable at Gd/Bi ratio ~ 0.01, which is much lower than the starting Gd/Bi ratio. Since the samples for EDS measurements are typical shiny single-crystal piece exfoliated from the ingot, we deduce that 1 atom% is a relatively stable doping level for Gd substitution at Bi sites in the aforementioned crystal growth process.

Before conducting electronic transport measurements on Gd:$Bi_2Se_3$ single crystals, we first revisit the electronic properties of $Bi_2Se_3$ single crystals grown with the same condition as that for Gd:$Bi_2Se_3$ crystals. The $Bi_2Se_3$ crystals possess a metallic ground state due to the Se vacancies. MR ($MR = [R(B) - R(0)]/R(0)$) in the 3–300 K and 0–14 T region are shown in Fig. 2(a). The total MR at low temperatures, e.g., below 30 K, are ~ 20% at 14 T, which slightly decrease to ~ 10% with heating. The observed roughly linear dependence of MR with the magnetic field is also reported in other 3D TIs [30,50]. Another interesting point is that the low temperature MR vs. magnetic field curves show obvious oscillation behaviors at high fields, which is due to the Landau quantization, namely, the Shubnikov-de Haas (SdH) oscillation. The pure oscillation patterns for 3 K ≤ $T$ ≤ 30 K can be obtained by subtracting the smooth MR backgrounds and are plotted against $1/B$ in Fig. 1(c). Note that, the background-subtracted oscillation patterns at different temperatures show exactly the same phases, however, with decreasing oscillation amplitude upon heating from 3 to 30 K. Therefore, the temperature-dependent SdH oscillation patterns are plotted in superposition [lower panel of Fig. 3(c)]. Since the $Bi_2Se_3$ crystals show metallic ground state, the SdH oscillations can be described by the Lifshitz-Kosevich (LK) formula, with the Berry phase being taken into account:

$$\frac{\Delta\rho}{\rho(0)} = \frac{5}{2}(\frac{B}{2F})^{\frac{1}{2}}R_T R_D R_S \cos(2\pi\left(\frac{F}{B}+\gamma-\delta\right))$$

Where $R_T = \alpha T\nu/B\sinh(\alpha T\mu/B)$, $R_D = \exp(-\alpha T_D \nu/B)$, and $R_S = \cos(\alpha g\nu/2)$. Here, $\nu = m^*/m_0$ is the ratio of the effective cyclotron mass $m^*$ to the free electron mass $m_0$; $g$ is the g-factor; $T_D$ is the Dingle temperature; and $\alpha = (2\pi^2 k_B m_0)/\hbar e$, where $k_B$ is Boltzmann constant, $\hbar$ is the reduced Planck constant, and $e$ is the elementary charge. The oscillation of $\Delta\rho$ is described by the cosine term with a phase factor $\gamma - \delta$, in which $\delta = 0$ for 2D Fermi pockets, and ±1/8 for 3D Fermi pockets, $\gamma = 1/2 - \Phi_B/2$, where $\Phi_B$ is the Berry phase. The Berry phase of SdH oscillations can also be obtained by extrapolating the Landau level index $n$ to the extreme field limit ($1/B \rightarrow 0$) in the Landau fan diagram, as shown in Fig. 2(d). Our resistivity measurements show that $\rho_{xx} \ll \rho_{xy}$ for $Bi_2Se_3$ crystals, we assign the maximum of the oscillations as half integer Landau indexes, the minimum of the oscillations as integer Landau indexes, respectively, and carefully fitted the data in straight lines [51,52]. As shown in Fig. 2(d), the intercept of the straight line with the y-axis is zero, meaning that the Fermi pocket contributing to SdH oscillations in our $Bi_2Se_3$ single crystals are topologically trivial bulk bands. Further, angular dependence of SdH oscillation measurements is conducted to analyze the morphology of the related Fermi pocket, as shown in Fig. 2(b). The angular dependent SdH oscillation measurements are plotted in Fig. 2(c), the FFT results of which are shown in Fig. 2(e). The angular dependent SdH oscillations show 2D-like behavior, which is probably contributed by some 2D-like trivial bands. Therefore, we deduce that the Fermi level is sitting on the edge of bulk conduction band, where the near-edge two-dimensional electron gas contributes mainly to the quantum oscillations [53-56]. The rest of the information obtained from the $Bi_2Se_3$ crystals together with the SdH of $Gd:Bi_2Se_3$ crystals will be discussed later.

Similarly, a freshly cleaved $Gd:Bi_2Se_3$ single crystal is employed to conduct MR measurements at various fixed temperatures, with applied external magnetic fields up to 14 T, as shown in Fig. 3(a). The quasi-linear MR with the magnetic field is similar to the observation in pure $Bi_2Se_3$ crystals, and the maximum MR values (e.g., MR at 14 T is ~ 8% below 50 K) decrease slightly with heating from 3 K. Another similarity is the low temperature oscillation patterns, which can be obtained by subtracting smooth backgrounds, as shown in Fig. 3(b). The SdH oscillation patterns are in phase with each other at different temperatures, however, the oscillation amplitudes decrease with heating, as described by the thermal damping factor in the LK formula. Let's pay attention to the Berry phase in $Gd:Bi_2Se_3$, obtained from the Landau fan diagram shown in Fig. 3(c). During linear fitting, we employ the same assigning rule as that for the $Bi_2Se_3$ crystals since $\rho_{xx} \ll \rho_{xy}$ also manifests here. The intercept of the straight line with the y-axis is 0.4, as shown in the zoom-in plot at high field limit [Inset of Fig. 3(c)]. Therefore, a non-zero Berry phase of 0.8 $\pi$ was obtained with 1 atom% Gd doping in the $Bi_2Se_3$. To further illustrate this point, we employ the LK formula to fit the oscillation patterns at 3 K, as shown in the inset of Fig. 3(b). A Berry phase of 0.78 $\pi$ was obtained from the fitting, which agrees well with that obtained from the Landau fan diagram. Further, we conducted magnetic measurements on the Gd:$Bi_2Se_3$ crystal to search for possible magnetic ordering which may contribute to the Berry phase. As shown in Fig. 3(d), the ZFC and FC curves coincide with each other perfectly and there's no obvious magnetic transition. The magnetism of trivalent lanthanide group ions originates from unpaired electrons, which can be described by the Curie law $\chi = C/T$ and Curie–Weiss law $\chi = C/(T + \theta_p)$. Here $C$ is the Curie constant, and the Weiss constant $\theta_p$ typically accounts for the magnetic ordering of the electronic moments below the Curie or Neel temperature for uncorrelated spins. The fitting of the Curie-Weiss law is shown in the inset of Fig. 3(d), the obtained Curie constant $C$ is ~ 2.5 emu K/mol and $\theta_p$ is ~ 0.4 K, which are neglectable. Here, $C=(N_A\mu_{eff}^2)/3k_B$, where $N_A$ is the Avogadro number, $k_B$ is the Boltzmann constant, therefore, the effective momentum $\mu_{eff}$ of Gd is ~ 4.5 $\mu_B$. The MH curves at 3, 30, and 300 K also show paramagnetic behaviors, as shown in Fig. 3(f). Therefore, the Gd dopants in the crystals show paramagnetism, indicating that the origin of this Berry phase is not long-range magnetic ordering.

Further, one may learn the fermiology from the SdH oscillations using the LK formula, e.g., the area of Fermi pockets, the Fermi vector, Fermi velocity, and the cyclotron mass of electron, etc. According to the Onsager-Lifshitz equation, the frequency of quantum oscillations, $F = (\varphi_0/2\pi^2)A_F$, where $A_F$ is the extremal cross-sectional area of the Fermi surface perpendicular to the magnetic field, and $\varphi_0$ is the magnetic flux quantum. The quantum oscillation frequencies of $Bi_2Se_3$ and $Gd:Bi_2Se_3$ from the fast Fourier transform results in Fig. 2(f) and 3(e), are 157 T and 151 T, respectively. Therefore, the cross-section area $A_F$ related to the Fermi pockets in $Bi_2Se_3$ and $Gd:Bi_2Se_3$ are 0.15 and 0.144 Å$^{-2}$, respectively. Using $A_F = \pi k_F^2$, the Fermi wave vector $k_F$ are 0.22 and 0.21 Å$^{-1}$ for the $Bi_2Se_3$ and $Gd:Bi_2Se_3$, respectively. According to the LK formula, the effective mass of carriers contributing to the SdH effect can be obtained through fitting the temperature dependence of the oscillation amplitude to the thermal damping factor $R_T$, which is shown in Fig. 4(a). During the fitting, we employ the normalized FFT amplitudes at various temperatures. The effective mass obtained from $R_T$ fitting are $0.2m_e$ for $Bi_2Se_3$ and $0.16m_e$ for $Gd:Bi_2Se_3$. Thus, one can obtain the Fermi velocity $v_F = \hbar k_F/m^* \sim 2.2 \times 10^5$ m/s for $Bi_2Se_3$ and $2.8 \times 10^5$ m/s for $Gd:Bi_2Se_3$. From the field damping relationship, we fit the Dingle temperatures via Dingle plot, as shown in Fig. 4(b), for both samples, which are 32 and 21 K for $Bi_2Se_3$ and $Gd:Bi_2Se_3$, respectively. The quantum relaxation time and quantum mobility can also be obtained by $\tau=\hbar/2\pi k_B T_D$ (3.8 and 5.8 fs for $Bi_2Se_3$ and $Gd:Bi_2Se_3$) and $\mu_Q=e\tau/m*$ (310 and 637 cm$^2$/Vs for $Bi_2Se_3$ and $Gd:Bi_2Se_3$), respectively. The carrier's parameters calculated from the quantum oscillations are summarized in Table 2, together with carrier's density and mobility. The Hall measurements are also conducted on the $Bi_2Se_3$ and $Gd:Bi_2Se_3$ single crystals from 3 to 300 K, and the carrier's properties calculated from which are plotted in Fig. 4(c). Further, we conduct the angular dependent SdH measurements on a $Gd:Bi_2Se_3$ single crystal at 3 K, as shown in Fig. 5(a). The quantum oscillation patterns are clear in low angle area, which is similar to those for $Bi_2Se_3$ [Fig. 2(c)]. The oscillation frequencies can be obtained via FFT spectra, as shown in Fig. 5(b). Interestingly, the frequency seems not shifting obviously during the rotation of the crystal, which is different from the 2D-like behavior in $Bi_2Se_3$. Therefore, we plot the FFT frequencies of $Bi_2Se_3$ and $Gd:Bi_2Se_3$ against the rotation angle in Fig. 5(c). Note that, for a 2D Fermi surface, the angular-dependent SdH frequency $F(\theta)$ increases in an inverse cosine rule: $F(\theta) = F(0)/\cos(\theta)$, indicated by the light blue area in Fig. 5(c). As expected, the $Bi_2Se_3$ follows 2D rotation rule during rotation. However, the SdH frequencies of $Gd:Bi_2Se_3$ are roughly a constant during rotation, which indicates that the related Fermi pocket is not contributed by 2DEG. Further, we employ Landau fan diagram to analyze the Berry phase shifting at different rotation angles, which are also shown in Fig. 5(c). The Berry phase for $Bi_2Se_3$ is always zero, since its oscillations are contributed by 2DEG. However, for the $Gd:Bi_2Se_3$, the Berry phase keeps changing during rotation, regardless of the nearly constant frequency. Therefore, the Gd doping has a limited influence on the morphology of the Fermi surface, however contributes to the band topology of dramatically.

## IV. Conclusions

We employed a modified Bridgeman method to grow Gd-doped $Bi_2Se_3$ single-crystal topological insulators. The Gd dopant distributes uniformly in the crystals, in which the doping level is ~1 atom% in Bi sites, as verified by EDS. The magnetic measurements show that no long-range magnetic ordering forms in $Gd:Bi_2Se_3$ crystals above 3 K, ruling out the possible magnetic ordering induced Berry phase. Magnetotransport measurements on $Gd:Bi_2Se_3$ single crystals show that Gd-doping has minor effects on SdH oscillation frequency, Fermi surface area, Fermi wave vectors, and effective electron mass of $Bi_2Se_3$ crystals. However, Gd-doping reduces the Hall mobility and carrier density, and modify the Berry phase and the morphology and topology of the Fermi pocket of $Bi_2Se_3$, namely, the major contribution of Fermi pocket changes from 2D-like electron gas states with trivial topology and zero Berry Phase to topological surface states with nontrivial topology and 0.8π Berry phase. The engineering of Berry phase by Gd doping may provide more possibility on application of topological insulators in electronic/spintronic devices.

## Acknowledgements

This work is supported by National Natural Science Foundation of China (Grant No. 11974155). WZ, AB, DC, XW, JK acknowledge the supporting from ARC Centre of Excellence in Future Low-Energy Electronics Technologies No. CE170100039.

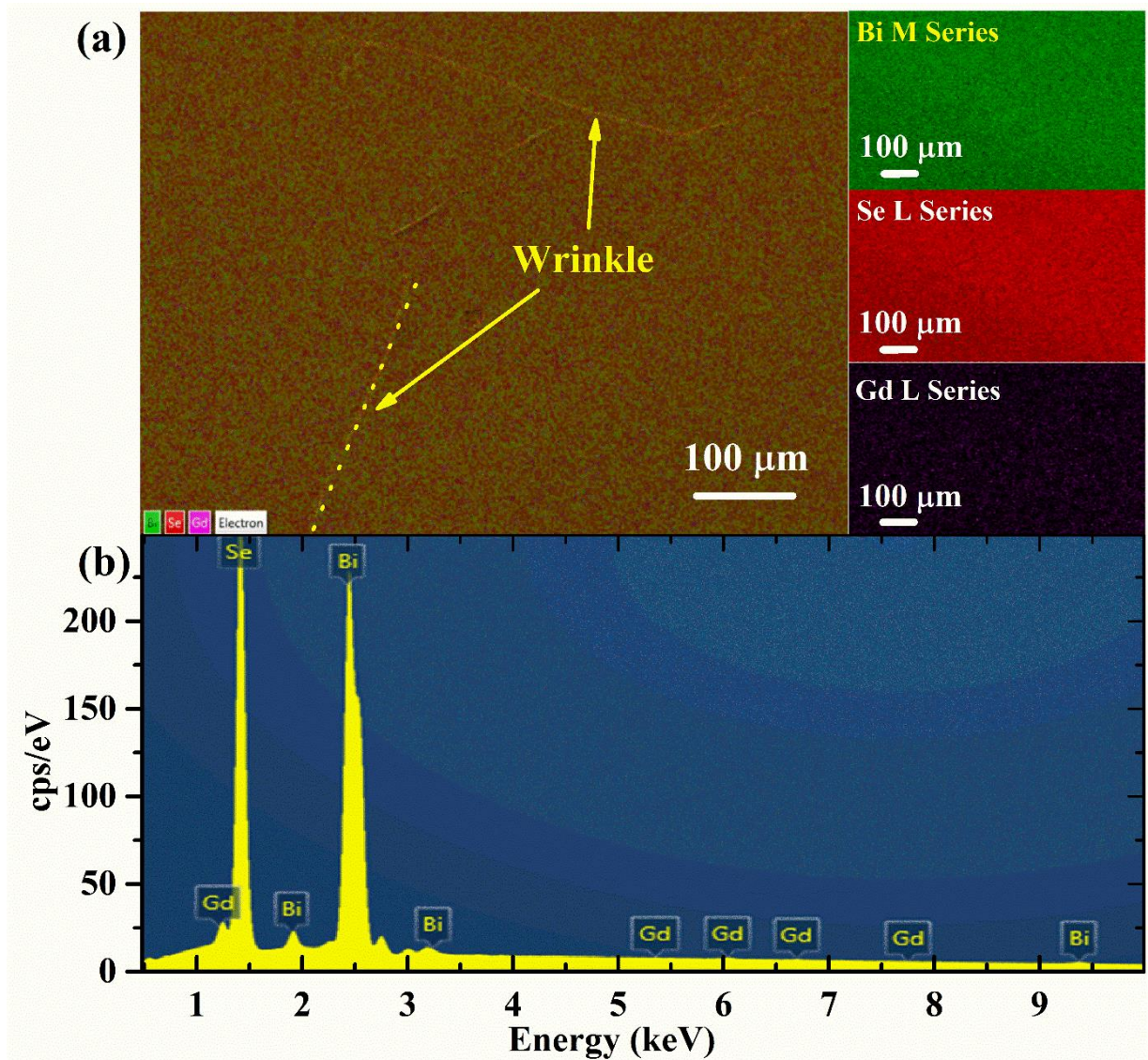

**Fig. 1.** SEM and EDS characterization of a freshly cleaved Gd:Bi$_2$Se$_3$ single-crystal surface. (a) The secondary electron image and elemental mapping image are plotted in superposition to illustrate the uniform distribution of all elements. The element mapping of Bi, Se and Gd in the same area are shown separately on the right side. (b) The energy dispersive spectroscopy of the area scan, in which the peaks are indexed with elements.

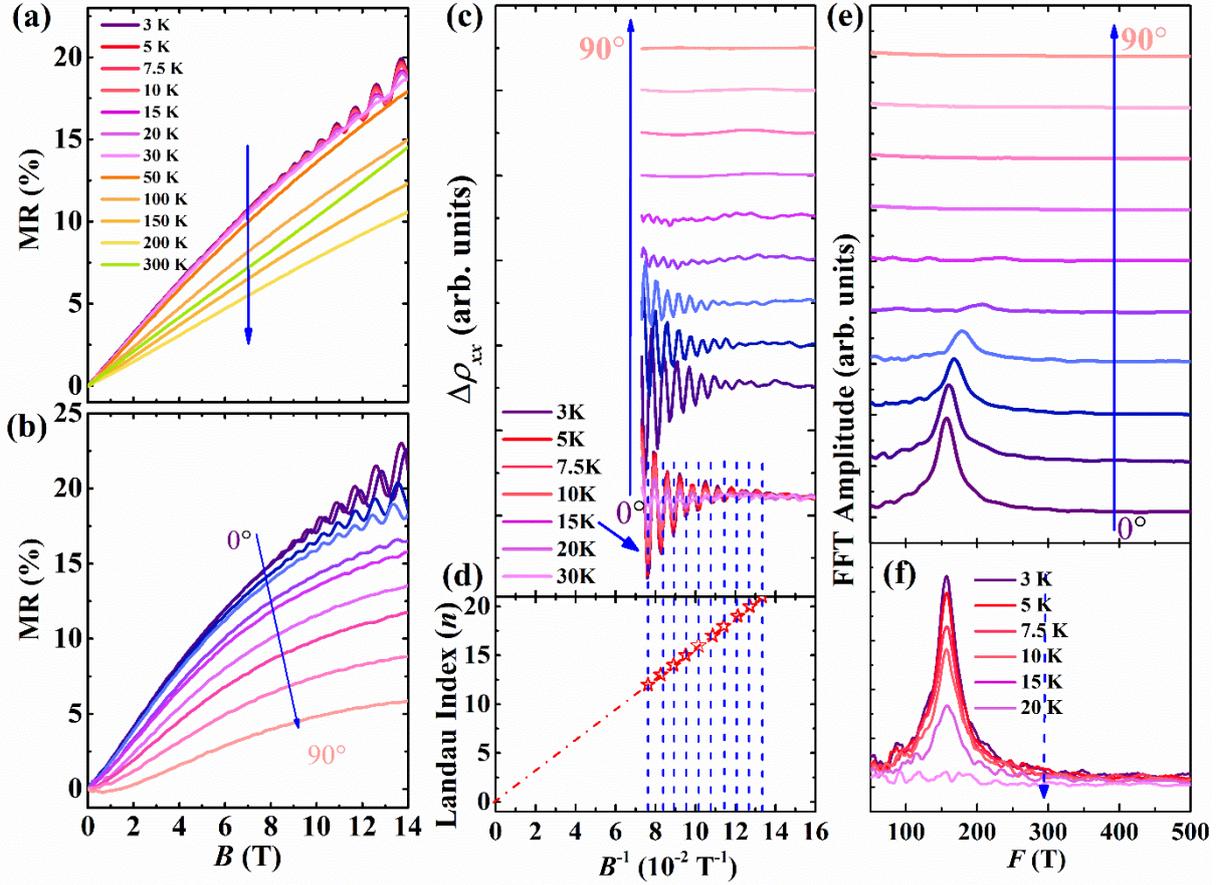

**Fig. 2.** A glimpse of the magnetotransport properties of a pure $Bi_2Se_3$ single crystal. (a, b) Temperature and angle ($T$=3 K) dependence of the MR vs. $B$ curves. (c) Temperature and angle dependent oscillation patterns whose amplitude decreases with heating (plotted in superposition of all temperatures, as pointed by the blue arrow) and increasing rotating angle. (d) The Landau fan diagram of the oscillation patterns at 0 degree. (e, f) FFT amplitude plot during rotating and heating.

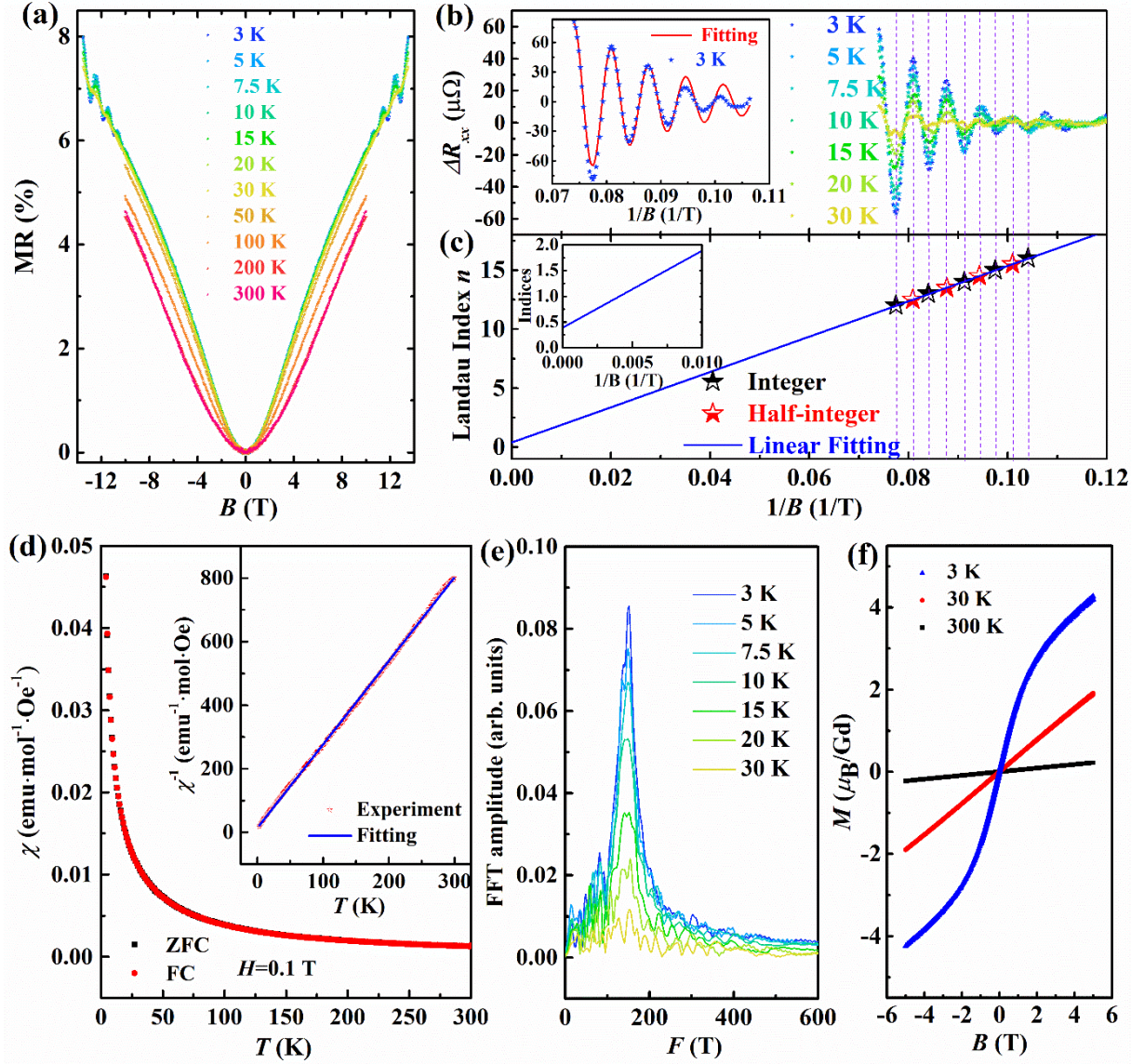

**Fig. 3.** A glimpse of the magnetotransport and magnetic properties of a Gd:Bi$_2$Se$_3$ single crystal. (a) MR vs. *B* curves at fixed temperatures ranging from 3 to 300 K. (b) The oscillatory patterns of the MR vs. 1/*B* curves. Inset: A fitting of the 3-K oscillatory pattern using the LK formula. (c) The Landau fan diagram is plotted to illustrate the Berry phase, in which the dot lines are guiding between the oscillation peaks and dips. (d) Temperature dependence of the ZFC and FC magnetic susceptibility. Inset: a fitting of the magnetic data using the Curie-Weiss law. (e) The FFT spectra of oscillatory patterns. (f) Magnetic hysteresis loops at *T*=3, 30, and 300 K.

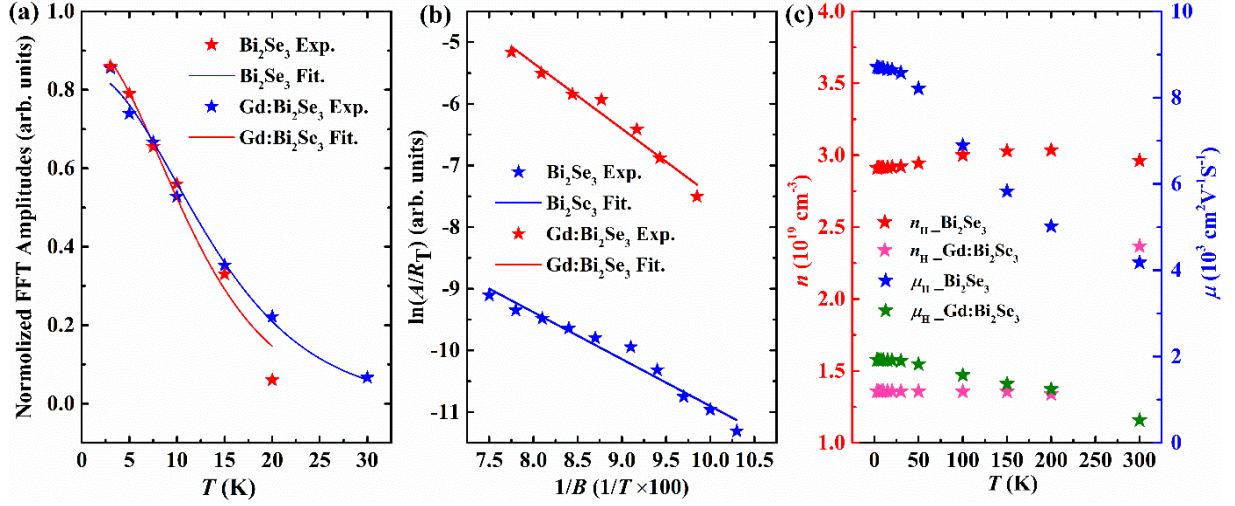

**Fig. 4.** The carrier's properties of $Bi_2Se_3$ and $Gd:Bi_2Se_3$ single crystals. (a) The LK formula fitting to obtain the effective mass of electrons. (b) The Dingle plots. (c) Temperature dependence of the carrier density and mobility obtained from Hall measurements.

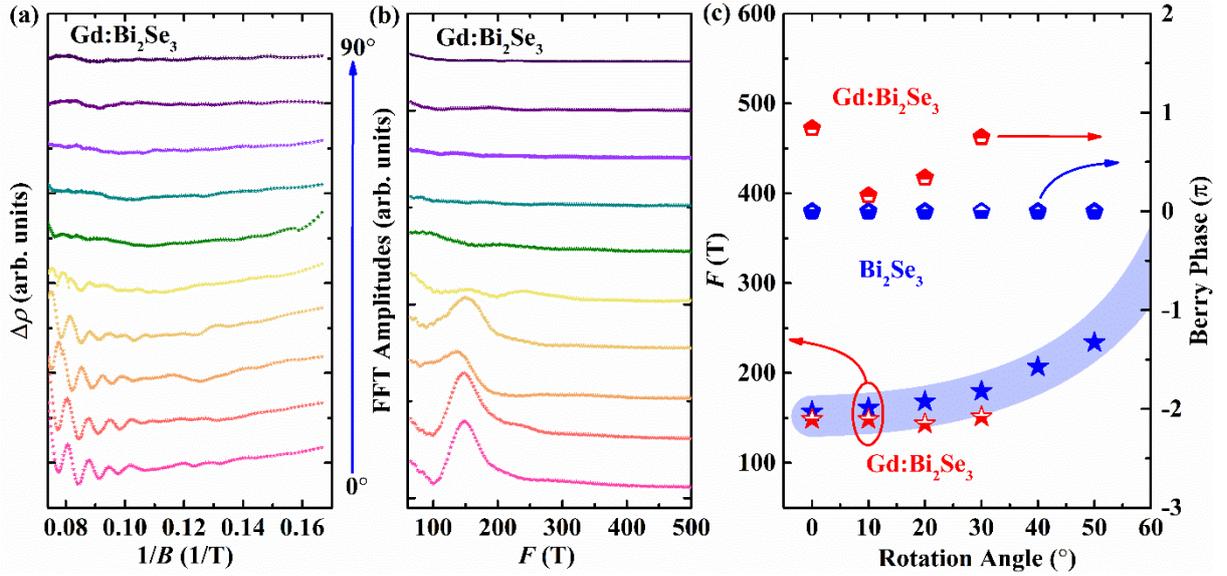

**Fig. 5.** Angular dependent SdH oscillations of $Gd:Bi_2Se_3$ single crystals. (a) The SdH oscillation patterns obtained from the angular dependent MR curves from which smooth backgrounds have been subtracted. (b) FFT spectra of the oscillation patterns in Panel (a). (c) The SdH oscillation frequencies and Berry phase are plotted as a function of the rotation angle. Note that, the light blue area indicates the frequency change of an ideal 2D Fermi surface.

**Table 1.** Elemental concentration of Gd:Bi$_2$Se$_3$ single crystals.

| Bi$_2$Se$_3$ crystals | Bi (atom%) | Se (atom%) | Gd (atom%) |
|---|---|---|---|
| Sample 1 | 40 | 59.6 | 0.4 |
| Sample 2 | 40.3 | 59.2 | 0.4 |

**Table 2.** The quantum oscillation and carrier's parameters at $T=3$ K: SdH frequency $F$, cross-sectional Fermi surface area $A_F$, Fermi vector $k_F$, effective electron mass $m^*$, Fermi velocity $v_F$, quantum mobility $\mu_Q$, carrier density ($n_{H\_3K}$) and mobility ($\mu_{H\_3K}$) from Hall measurements, and the Berry phase $\Phi_B$.

| | $F$ (T) | $A_F$ (Å$^{-2}$) | $k_F$ (Å$^{-1}$) | $m^*$ ($m_e$) | $v_F$ (m/s) | $\mu_Q$ (cm$^2$/Vs) | $n_{H\_3K}$ (cm$^{-3}$) | $\mu_{H\_3K}$ (cm$^2$/Vs) | $\Phi_B$ |
|---|---|---|---|---|---|---|---|---|---|
| Bi$_2$Se$_3$ | 157 | 0.15 | 0.22 | 0.2 | 2.2×10$^5$ | 310 | 2.8×10$^{19}$ | 8400 | 0 |
| Gd:Bi$_2$Se$_3$ | 151 | 0.14 | 0.21 | 0.16 | 2.8×10$^5$ | 637 | 1.38×10$^{19}$ | 1500 | 0.8π |